\documentclass[12pt]{article}

\usepackage{graphicx}
\usepackage{amsmath}
\usepackage{amssymb}
\usepackage{bm}
\usepackage{graphics}

\begin{document}  
 
 \title{Optical properties of metal nanoparticles with arbitrary shapes} 

 \author{Iv\'an O. Sosa\thanks{E-mail: ivan@fisica.unam.mx}, Cecilia Noguez\thanks{Author to whom all correspondence should be addressed.   E-mail: cecilia@fisica.unam.mx}~~and Rub\'en G. Barrera\thanks{E-mail: rbarrera@fisica.unam.mx}\\
 \it Instituto de F\'{\i}sica, Universidad Nacional Aut\'onoma
 de M\'exico, \\ Apartado Postal 20-364, Distrito Federal 01000,  M\'exico}

 \maketitle 

\begin{abstract}
We have studied the optical properties of metallic nanoparticles with arbitrary shape. We performed theoretical calculations of the absorption, extinction and scattering efficiencies, which can be directly compared with experiments, using the Discrete Dipole Approximation (DDA). In this work, the main features in the optical spectra have been investigated depending of the geometry and size of the nanoparticles. The origin of the optical spectra are discussed in terms of the size, shape and material properties of each nanoparticle, showing that a nanoparticle can be
distinguish by its optical signature.  
\end{abstract}

\newpage

 
\section{Introduction}
The fabrication of nanostructures requires a deep understanding
of the physical phenomena involved at this length-scale. Low-dimensional quantum structures have shown to have unique optical and electronic properties. In particular, the shape and size of low-dimensional structures are crucial parameters that determine those physical properties. The characterization of these parameters are important issues either in fundamental research or in technological applications, covering from growth and characterization to device processing. Among nanostructures, metallic nanoparticles are also important because some of their main physical properties might be completely different from the corresponding ones in either molecules or bulk solids. For example, they might assume crystal structures that do not correspond to those of the bulk solid.~\cite{opto,yaca}. Also, the catalytic activity of some of them depends strongly on their size and shape.

Typically, the estimation of the shape and size of nanoparticles
has been done using techniques like Atomic Force Microscopy (AFM),
Scanning Tunneling Microscopy (STM), Transmission Electron Microscopy (TEM) and Reflection High-Energy Electron Diffraction (RHEED)~\cite{yaca,dami,zheng,kim,cirlin}. These techniques provide the image of a small  piece of the sample, this means that they give information about local properties by characterizing a few nanoparticles at a time. Different shapes of nanoparticles have been reported with the use of these techniques, such as spheres, spheroids, lens-shaped, cone-shaped, pyramids with different facets, truncated pyramids, and different types of polyhedra~\cite{liu,zou,yang,yaca}. A vast amount of information has been obtained through these ``structural-characterization techniques'', however, they still have some limitations. One limitation is that in most cases the growth and characterization are made in different ambients, which is a serious problem because the properties of nanoparticles are ambient dependent. In some of these techniques the sample is literally touched during characterization and sometimes this might substantially modify the properties of a nanoparticle. Furthermore, the growth and characterization of nanoparticles are usually made at different times, and this might become an additional uncontrollable variable. 

The limitations of these ``structural-characterization techniques'' make
desirable the use of complementary tools that could accomplish the same
objectives but in the same ambient, in real time, in a non-destructive way, and providing statistical properties of the whole sample. Among
these complementary tools one finds that within this context optical  spectroscopies have been extremely useful  due to their non-destructive and real-time character together with their in situ potentiality~\cite{roman,beitia,roman2}. The above attributes can be found, for example, in Light Absorption Spectroscopy, Surface Enhanced Raman Scattering (SERS), Differential Reflectance (DR), etc.~\cite{roman,beitia,roman2}. These attributes  have allowed to control the growth of superlattices~\cite{dr}, and with a proper implementation of them, it might be possible to control also the growth of nanoparticles, correcting their shape and size during the growth process. In the future, this fact will be of crucial importance for the development of nanosciences and their technological applications.

A variety of results using optical techniques have been able to relate the
spectroscopic features of the nanoparticles spectra with the excitation of surface plasmons and excitons, as well as the significant enhancement in Raman-absorption
-excitation peak intensities, with the size and shape of a given system~\cite{roman,nordin,yang2,ceciMRS}. These features are manifested in optical spectra such as in light absorption, reflection and transmission or in light-extinction and scattering spectra, however, a clear physical interpretation of this spectroscopic information is still waiting. In conclusion, the actual correct determination of the size and shape parameters of a given nanoparticle is still controversial because a more complete experimental determination is needed, together with a corresponding clear physical interpretation.

In the present work, we study the optical properties of metallic nanometer-sized particles with different shapes and sizes. We calculate and discuss spectra for the extinction, absorption and scattering efficiencies. In particular, we show results for gold and silver nanoparticles. Our main goal is twofold, first we show how the main peaks of the optical spectra can be associated to the shape and size of the nanoparticle, as well as its material properties, and second we show the relative importance of absorption and scattering processes as a function of the same geometrical and material parameters. Although our calculation procedure is, at
the present time, unable to display the actual fields induced in the particle, we can attach a multipolar character to the different excitations responsible for the charateristic features of the spectra by comparing results for particles with the same shape but with different size. We have also paid a particular attention to the accuracy of our calculations showing explicitly the conditions under which they might become numerically unstable, and comparing them with similar calculations performed by other authors. We believe that this study can be very useful to determine and
optimize some of the physical properties of nanoparticles by controlling their shape and size during and after a growth process. Another goal is to motivate future measurements of extinction and absorption spectra of nanoparticles.

\section{Formalism}
In this work, the nanoparticles of interest are typically large enough to accurately apply classical electromagnetic theory to describe their interaction with light \cite{lance}. But they are also small enough to observe strong variations in the optical properties depending on the particle size, shape, and local environment. Because of the complexity of the systems being studied, efficient computational methods capable of treating large size materials are essential. In the last few
years, several numerical methods have been developed to determine the optical properties of small particles, such as the Discrete Dipole Approximation (DDA), T-matrix  and Spectral Representation methods (SR) \cite{mish}.

In this work we employed the DDA, which is a computational procedure suitable for studying scattering and absorption of electromagnetic radiation by particles with sizes of the order or less of the wavelength of the incident light. DDA has been applied to a broad range of problems, including interstellar and interplanetary dust
grains, ice crystals in the atmosphere, human blood cells, surface features of semiconductors, metal nanoparticles and their aggregates, and more. The DDA was first introduced by Purcell and Pennypacker \cite{purcell}, and has been subjected to several improvements, in particular those made by Draine, and collaborators~\cite{draine}. Below, we briefly describe the main characteristics of DDA and its numerical implementation: the DDSCAT code. For a full description of DDA and DDSCAT, the reader may consult Refs.~[21-23].

The main idea behind DDA is to approximate a scatterer, in our case the nanoparticle, by an large enough array of polarizable point dipoles. Once the location and polarizability of each dipole are specified, the calculation of the scattering and absorption efficiencies by the dipole array can be performed, depending only on the accuracy of the mathematical algorithms and the capabilities of the computational hardware. Although the calculation of the radiated fields in DDA is, in principle,
also possible, it is actually beyond the computational capabilities of any of the present systems.

\subsection{Discrete Dipole Approximation}

Let us assume an array of $N$ polarizable point dipoles located at $ \{\mathbf{r}_{i}\}, \, \, i=1,2,\dots ,N$, each one characterized by a polarizability  $\alpha_{i}$. The system is excited by a monochromatic incident plane wave $\mathbf{E}_{\mathrm{inc}}(\mathbf{r},t) = \mathbf{E}_{\mathrm{0}}e^{i\mathbf{k\cdot r}-i\omega t}$, where $\mathbf{r}$ is the position vector, $t$ is time, $\omega $ is the angular frequency, $k=\omega /c=2\pi/\lambda $ is the wavevector, $c$ is the speed of light, and $\lambda $ is the wavelength of the incident light. Each dipole of the system is subjected to an electric field that can be split in two contributions: (i) the incident radiation field, plus (ii) the field radiated by all the other induced dipoles. The sum of both fields is the so called local field at each dipole and is given by 
\begin{equation}
\mathbf{E}_{i,\mathrm{loc}}=\mathbf{E}_{i,\mathrm{inc}}+\mathbf{E}_{i,\mathrm{dip}}=\mathbf{E}_{\mathrm{0}}e^{i\mathbf{k\cdot r_i}}-\sum_{i\neq j}\mathbf{A}_{ij}\cdot \mathbf{P}_{j},
\end{equation}
where $\mathbf{P}_{i}$ is the dipole moment of the $i$th-element, and $\mathbf{A}_{ij}$ with $i \neq j$ is an interaction matrix with $3\times 3$ matrices as elements, such that 
\begin{eqnarray}
\mathbf{A}_{ij}\cdot \mathbf{P}_{j} &=&\frac{e^{ikr_{ij}}}{r_{ij}^{3}}\{k^{2}\mathbf{r}_{ij}\times (\mathbf{r}_{ij}\times \mathbf{P}_{j}) \\
&+&\frac{(1-ikr_{ij})}{r_{ij}^{2}}\left[ r_{ij}^{2}\mathbf{P}_{j}-3\mathbf{r}_{ij}(\mathbf{r}_{ij}\cdot \mathbf{P}_{j})\right] \}.  
\end{eqnarray}
Here $r_{ij}=|\mathbf{r}_{i}-\mathbf{r}_{j}|$, and $\mathbf{r}_{ij} =\mathbf{r}_{i} - \mathbf{r}_{j}$, and we are using cgs units. Once we solve the $3N$-coupled complex linear equations given by relation 
\begin{equation}
\mathbf{P}_{i}=\alpha_{i}\cdot \mathbf{E}_{i,\mathrm{loc}},
\end{equation}
and determined each dipole moment $\mathbf{P}_{i}$, we can then find the extinction and absorption cross sections for a target, $C_{\mathrm{ext}}$ and $C_{\mathrm{abs}}$, in terms of the dipole moments as  
\begin{eqnarray}
&&C_{\mathrm{ext}}=\frac{4\pi k}{|\mathbf{E}_{0}|^{2}}\sum_{i=1}^{N}\mathrm{Im}(\mathbf{E}_{i,\mathrm{inc}}^{\ast }\cdot \mathbf{P}_{i}) \\
&&C_{\mathrm{abs}}=\frac{4\pi k}{|\mathbf{E}_{0}|^{2}}\sum_{i=1}^{N}\{\mathrm{Im}[\mathbf{P}_{i}\cdot (\boldsymbol{\alpha}_{i}^{-1})^{\ast }\mathbf{P}_{i}^{\ast }]-\frac{2}{3}k^{3}|\mathbf{P}_{i}|^{2}\},\,\,\,
\end{eqnarray}
where $\ast $ means complex conjugate. The scattering cross section can be obtained using the following relation,  
\begin{equation}
C_{\mathrm{ext}}=C_{\mathrm{sca}}+C_{\mathrm{abs}}.
\end{equation}

There is some arbitrariness in the construction of the array of dipole points that represent a solid target of a given geometry. For example, the geometry of the grid where the dipoles have to be located is not uniquely determined and a cubic grid is usually chosen. Also, it is not obvious how many dipoles are required to adequately approximate the target, or which is the best choice of the dipole polarizabilities. If one chooses the separation between dipoles $d$ such that $d<<\lambda $, then, one can assign the polarizability for each particle $i$ in vacuum, using the Lattice Dispersion Relation (LDR) polarizability, $\alpha_{i}^{\rm LDR}$ at a third order in $k$, given by~\cite{draine}
\begin{equation}
\alpha_{i}^{\rm LDR}=\frac{\alpha_{i}^{\rm CM}}{1 + \alpha_{i}^{\rm CM}\left[b_1+ b_2 \boldsymbol{\epsilon}_{i}+ b_3 S \boldsymbol{\epsilon}_{i} \right](k^2/d)},
\end{equation}
where $\epsilon_{i}$ is the macroscopic dielectric function of the particle, $\alpha_{i}^{\rm CM}$ is the polarizability given by the well known Clausius-Mossotti relation,  
\begin{equation}
\alpha_{i}=\left(\frac{d}{3}\right)^3 \frac{(\epsilon_{i} - 1)}{(\epsilon_{i}+2)},
\end{equation}
and $S$, $b_1$, $b_2$, and $b_3$ are coefficients of the expansion.

Now the question is, how many dipoles we need to mimic the continuum macroscopic particle with an array of discrete dipoles? The answer is not straightforward, since we have to consider the convergence of the physical quantities as a function of the dipole number. It has been found that $N\geq 10^{4}$ for an arbitrary geometry is a good starting number, as shown in the Appendix. However, we have a matrix of $(3N)^{2}$ complex elements which would require a large amount of computational effort.

In this work we have employed the code adapted by Draine and Flatau to solve the complex linear equations found in DDA. To solve the complex linear equations directly would require tremendous computer capabilities, however, one can use iterative techniques to compute the vector $\mathbf{P\equiv }\{\mathbf{P}_{i}\}$. In this case, each iteration involves the evaluation of matrix-vector products such as $\mathbf{A}\cdot \mathbf{P}^{\mathrm{(n)}}$, where $n$ is the number of the iteration. The algorithm, named DDSCAT, locates the dipoles in a periodic cubic lattice, and
it is possible to use fast Fourier transform techniques to evaluate  matrix-vector products such as $\mathbf{A}\cdot \mathbf{P}$, which allows the whole computation of the final vector $\mathbf{P}$ for a large number of dipoles \cite{ddscat}. For a detailed description of DDA and DDSCAT code, the reader can look at Refs.~[22-25].

\section{Results and Discussion}

We define the extinction, absorption and scattering efficiencies or
coefficients, $Q_{\mathrm{ext}}$, $Q_{\mathrm{sca}}$ and $Q_{\mathrm{abs}}$ 
as, 
\begin{equation}
Q_{\mathrm{ext}}=\frac{C_{\mathrm{ext}}}{A},\quad Q_{\mathrm{abs}}=\frac{C_{%
\mathrm{abs}}}{A},\quad Q_{\mathrm{sca}}=\frac{C_{\mathrm{sca}}}{A}.
\end{equation}%
where $A=\pi a_{\mathrm{eff}}^{2}$, and $a_{\mathrm{eff}}$ 
is defined through the concept of an effective volume equal to $4\pi a_{\mathrm{eff}}^{3}/3.$ In Figs. 1 to 8 we show $Q_{\mathrm{ext}}$, $Q_{\mathrm{sca}}$ and $Q_{\mathrm{abs}}$ in dotted-, dashed- and
solid-lines, respectively, as a function of the wavelength of the incident
light, $\lambda $, for nanometric-size particles. All the calculations were done for nanoparticles  with $a_{\mathrm{eff}}$ $=$50~nm, and dielectric functions as measured on bulk silver and gold by Johnson and Christy~\cite{JandC}. The nanoparticles were represented or mimic by around 65,000 point-dipoles in order to have a good convergence on their optical properties, as we will discuss below. This number of dipoles is quite large in comparison with the numbers used in previous studies on isolated and supported small metallic nanoparticles~\cite{yang2,otros} where, incidentally, only the extinction efficiency was reported. Our use of a large number of dipoles is in agreement with the results found in a  a previous work where
we found that even for small metallic nanoparticles with radii of about a few nanometers, one needed more than 12,000 dipoles to achieve convergence on their optical properties~\cite{ceciMRS}. In particular, we found that the extinction efficiency converges very rapidly as a function of the number of dipoles. However, this is not the case for the absorption or scattering efficiencies where a very large number of dipoles were necessary to achieve such a convergence. As we show in this paper, to elucidate the optical properties of large nanoparticles it is indispensable to do an in-depth study of the scattering and absorption efficiencies, and not only of the extinction one.

In Fig.~1, we show the optical efficiencies for a sphere of radius of 50~nm.
In the spectra we can observe that at about 320~nm all the efficiencies have a local minimum that corresponds to the wavelength at which the dielectric function of silver, both real and imaginary parts, almost vanish. Therefore, this feature of the spectra is inherent to the material properties and, as we observe below, it is independent of the particle geometry. Below 320~nm, the absorption of light is mainly due to the intra-band electronic transitions of silver, therefore, this feature of the spectra 
should be also quite independent of the shape and size of the particles, as it is actually corroborated in all the graphs shown below corresponding to a silver particle. At about 350~nm, the spectrum of $Q_{\mathrm{abs}}$ shows a peak that is related to the excitation of the surface plasmon of the sphere, therefore this feature is inherent to the geometry of the particle, although the position depends on 
its material properties. At larger wavelengths  the spectrum of $Q_{\mathrm{abs}}$ shows  specific features from 380~nm to about 500~nm that corresponds to plasmon excitations due to higher multipolar charge distributions~\cite{multipoles}. If we look only at the $Q_{\mathrm{ext}}$ it is not possible to observe such  features since scattering effects hide them. The $Q_{\mathrm{sca}}$ spectrum shows a broad structure from 320~nm to 750~nm,  with a maximum at about 400~nm, which is three times more intense than the maximum of $Q_{\mathrm{abs}}$. The characteristics of the $Q_{\mathrm{sca}}$ spectrum are mainly due to the  size of the particle. In a previous work~\cite{ceciMRS} we found that this maximum is less pronounced as the radius of the sphere  increases. Also, the $Q_{\mathrm{sca}}$ spectra of nanospheres decay slowly as the radius  increases; this only means that as the sphere becomes larger it scatters light at  longer wavelengths, as expected.

In Fig.~2, we show the optical efficiencies for a nanocube with sides of 83~nm. For wavelengths between 320~nm and 450~nm the main contributions to the $Q_{ \mathrm{ext}}$ spectrum comes from light absorption and scattering, while for 400~nm $<\lambda <$ 700~nm, it comes mainly from light scattering effects. The $Q_{\mathrm{abs}}$ spectrum shows a rich structure for $\lambda$ between 320~nm and 420~nm, where several peaks are observed. These peaks are  associated to the resonances inherent to the cubic geometry~\cite{fuchs}. The peak at $\lambda =410$~nm corresponds to the dipolar resonance, while the peaks at smaller wavelengths are due to high-multipolar excitations. From 420~nm to about 700~nm, $Q_{\mathrm{abs}}$ shows
a tail that corresponds also to high-multipolar excitations, however, these excitations are due to size effects. Therefore, this tail is observed in all figures since all the nanoparticles have the same volume. We also performed calculations for silver cubes with sides of 9~nm, where we found that for $\lambda <$ 410~nm  the $Q_{\mathrm{abs}}$  spectrum has more or less the same structure and intensity than the one shown here, but for larger wavelengths it does not  have a tail. Above the resonances, at about $\lambda =$450~nm, the $Q_{\mathrm{sca}}$ spectrum shows a maximum. We can observe that this maximum is at a larger wavelength than  the one corresponding to a sphere. This shift of the scattering peak to larger wavelengths is due to the increase in the size of the nanoparticle. The latter observation was also  described in a previous work~\cite{ceciMRS}, where the $Q_{\mathrm{ext}}$ and $Q_{\mathrm{abs}}$ spectra for nanocubes of $a_{\mathrm{eff}}=$50~nm  and $a_{\mathrm{eff}}=$ 150~nm, were compared. In Fig.~8, we will  display the optical efficiencies of a gold cube of the same size to  discuss the effects due to
material properties.

In Fig.~3 we show the optical efficiencies of a silver prolate spheroidal
nanoparticle where the incident electromagnetic field is taken parallel to
its minor axis. The spheroid has a major to minor axis ratio of 3 to 1. 
From 320~nm to about 450~nm  one can see that  the
contributions to the $Q_{\mathrm{ext}}$  spectrum from absorption
and scattering processes are more or less of the same order of magnitude.
Notice that the peak at  shorter wavelengths ($\sim 350$~nm) is
almost twice as big as the peak at  longer wavelengths ($\sim 430$%
~nm). The $Q_{\mathrm{abs}}$ spectrum is characterized by two peaks, one at $%
\sim 350$~nm, and the other at $\sim 430$~nm, and a tail from 500~nm to
750~nm. The first peak corresponds to the main resonance of an ellipsoid due
to a dipolar charge distribution and its position depends on the specific
geometry of the particle. On the other hand, the peak and the tail at 
 longer wavelengths correspond to excitations of higher multipolar
charge distributions. We recall that  although the DDA   %
cannot assign a specific multipolar order to those excitations~\cite%
{multipoles},  one can assert that in this particular geometry, the
high multipolar excitations are due to size effects  because they
are not observed  in small spheroidal nanoparticles. Furthermore, it
is necessary to do an in-depth study of both absorption and scattering
effects to clearly elucidate the problem. For example, it is known that
scattering effects of smaller particles have a maximum at smaller
wavelengths, therefore it is possible to hide the multipolar effects if they
appear at the same wavelength as scattering effects do. In such case, a
detail study of the absorption and extinction efficiencies is necessary. We
have also calculated the optical efficiencies for smaller ellipsoidal
nanoparticles with a major semiaxis of 12~nm. In this case, the main
contribution to $Q_{\mathrm{ext}}$ comes from light absorption processes due
to the excitation of one surface plasmon (dipolar excitation) whose location
depends on the particular geometry of each nanoparticle and on its material
properties. From 475~nm to 750~nm the main contribution to $Q_{\mathrm{ext}}$
come from light-scattering effects, although, we also observed a tail in $Q_{%
\mathrm{abs}}$.  Although this tail shows  also a few peaks, 
 they are washed out as the number of dipoles in the calculation is
increased dramatically~\cite{ceciMRS},  so they should come from
lack of convergence in the calculations.  It is interesting to note
that this lack of convergence is particularly observed in metal particles
and it could be due to the fact that at  long wavelengths the
dielectric function of silver, and in general of metals, is negative and
large~\cite{drude}.

In Fig. 4 we show the optical efficiencies for the same silver spheroidal
nanoparticle but  now with the incident electromagnetic field
perpendicular to its minor axis. Like for the previous orientation,  %
the spectrum of $Q_{\mathrm{abs}}$ shows also two peaks and a tail, however
the peak at 350~nm is less intense than the peak at about 430~nm. In this
case, the dipolar excitation or main surface plasmon is observed at $\lambda
=430$~nm, while high-multipolar excitations are observed at $\lambda =350$%
~nm.  These changes in the excitation energy of the multipolar modes
depending on the orientation of  the polarization of the incident
light have been  already calculated  for oblate and prolate
spheroids using  the exact solution in the  quasi-static
limit, where multipolar excitations are induced by the presence of a
substrate~\cite{roman2}.  Here the $Q_{\mathrm{ext}}$ spectrum has
again contributions from both absorption and scattering processes. From
475~nm to 750~nm the main contribution to $Q_{\mathrm{ext}}$ comes from
light scattering effects, as it does in Fig.~3. However, for this
orientation of the ellipsoid $Q_{\mathrm{ext}}$ is about twice  as
intense as the one corresponding to an incident electromagnetic
field parallel to the major axis. Besides that, it is possible to assign a
particular geometry  to each spectra, so we can also obtain
information about its orientation. However, since the intensity and
sharpness of the spectra is dominated by the material properties, it is
possible that this kind of information could be hidden if the dielectric
response of the particle is different~\cite{ceciMRS}. 

In Fig.~5 we show the optical efficiencies for a silver cylinder when the
incident electromagnetic field is parallel to its symmetry axis. The
cylinder has a radius of 43.7~nm and its axis is two times larger than 
 its radius. From 320~nm to 470~nm the absorption effects are as
important as the light scattering effects. The $Q_{\mathrm{abs}}$ spectrum
shows three main peaks which are due to the geometrical properties of the
nanoparticle, as we have  discussed before. The $Q_{\mathrm{abs}}$
spectrum also shows a tail at larger wavelengths due to high-multipolar
excitations coming from size effects. In $Q_{\mathrm{abs}}$ the peak at $%
\lambda \sim $430~nm corresponds to a dipolar excitations, while the other
two peaks at lower wavelengths are due to multipolar  excitations
inherent  to the geometry of the particle. These multipolar
excitations are enhanced due to size effects, since for smaller
nanocylinders we  observe that the dipolar excitation is fourteen
times more intense than other multipolar excitations. At  longer
wavelengths the spectrum is dominated by scattering effects, as shown in the
figure. For this geometry, we observe that the scattering processes are very
efficient as compared with  the corresponding ones in other
particles, and the main peak in the extinction spectrum is located at 
 longer wavelengths. This geometry is similar to the spheroidal
nanoparticles, therefore, we observed that both spectra are also similar,
except that the absorption of the cylinder shows an  additional
resonance.  

In Fig.~6 we show the optical efficiencies  of a silver tetrahedral
nanoparticle  when the incident  electric field is parallel
to its basis, and it is pointing  towards one of its vertex. The
tetrahedron has sides of 164~nm. The spectra for $\lambda >320$~nm are
different  from those discussed above, since a more complex
resonance structure is found. In this case, light absorption and scattering
occur over a wide range of wavelengths. For $\lambda <500$~nm the absorption
effects are a little more intense than scattering effects, and for $\lambda
>500$~nm the contrary occurs. For this particular geometry, we observe that
the spectra have a very rich structure coming from the excitation of surface
plasmons as well as from light scattering. The $Q_{\mathrm{abs}}$ spectrum
has a dipolar resonance at about 500~nm which is enhanced by multipolar
excitations from both geometry and size effects. At smaller wavelengths, the 
$Q_{\mathrm{abs}}$ spectrum is dominated by high-multipolar excitations due%
 \ to the geometry of the particle. These high-multipolar
excitations have been experimentally observed and theoretically identified
for Ag prisms~\cite{science,new}. The $Q_{\mathrm{sca}}$ spectrum shows a
maximum at about $\lambda =590$~nm. Notice that from 600~nm to 750~nm the $%
Q_{\mathrm{abs}}$ spectrum shows  several peaks  coming from
lack of convergence in the calculations, thus they have no physical meaning
as it was pointed out before.

In Fig.~7 we show the optical efficiencies for a silver nanoparticle that is
made of small spheres which are so arranged as to resemble a kind of pyramid 
 with a square base. The sides of the  base of the pyramid 
 is 146~nm while its high  is 73~nm. We have taken the
incident  electric field parallel to its  base. Like for the
tetrahedron, the spectra are completely different  from those
corresponding to the other silver nanoparticles discussed above. When $%
\lambda >320$~nm light absorption and scattering occur over a wide range of
wavelengths, being the absorption effects the ones that dominate the
structure of the spectrum. The spectra of the pyramid also have a very rich
structure coming from the excitation of surface plasmons as well as from the
light scattering corresponding to this particular geometry. However, the
structure of the $Q_{\mathrm{ext}}$  is rather smooth as compared to
the one of the tetrahedral nanoparticle. This could be associated to the
fact that the pyramid is constructed  by using small spheres, and
each sphere has a single surface plasmon and a  well defined
scattering peak. However, in this case it was not possible to distinguish
between dipolar and high-multipolar excitations. Again, we observe four
sharp peaks at large wavelengths  which come from lack of
convergence in the calculations at  those specific wavelengths,
nevertheless,  the spectra should be valid at all the other
wavelengths. The enhancement of these peaks could come from the fact that
the small spheres are touching, however a more careful study has to be done.

In Fig.~8 we show the optical efficiencies for a nanocube  with the
same volume as the one in Fig.~2, but  made out of gold. For the
gold cube the $Q_{\mathrm{ext}}$ spectrum shows a broad structure from
250~nm to about 500~nm, due mainly to absorption effects although light
scattering is also observed. Contrary to the rich structure found for the
silver cube, in this case we observe very smooth curves. This behavior 
 comes from the fact that the relaxation time of  electrons
in gold is about ten times smaller than the relaxation time for silver,
this fact  gives rise to a completely different dielectric response.
The several peaks found for the silver nanoparticles are now replaced by a
wide structure that is due to surface plasmons  as well as intra-band
transitions. Then, we can conclude that both effects are of the same order
of magnitude since the fine structure inherent to a cube has been washed
out. The latter gives rise to a wide peak in $Q_{\mathrm{abs}}$ at about
550~nm and a peak in $Q_{\mathrm{sca}}$ at about 580~nm. Notice that at $%
\lambda =580$~nm the $Q_{\mathrm{sca}}$  spectra of the silver
(Fig.~2) and gold (Fig.~8) nanocubes are of the same magnitude. However for
the silver cube the $Q_{\mathrm{sca}}$ has its maximum at lower wavelengths
and it is almost twice as large as the one corresponding to the
gold cube, where its maximum  seems to be inhibited and  this
might be due to intra-band transitions.

Recently silver and gold suspended nanoparticles have been produced by wet
chemistry techniques~\cite{science} and by laser ablation in aqueous
solution~\cite{mafune1,mafune2}. In the first case, the nanoparticles have
been fully characterized  by measurements of the extinction
efficiency~\cite{science}. In the  second case,  the authors reported  the production of gold nanoparticles with a diameter
between 1~nm to 50~nm, as well as silver nanoparticles with diameters
between 1~nm to 40~nm. The reported absorption spectrum  of silver
nanoparticles show a minimum at about 320~nm, and a wide peak with a maximum
around 400~nm, and at about 600~nm the spectrum  starts to decay 
very fast. On the other hand, for gold nanoparticles the absorption spectrum
shows a wider structure from 300~nm to 800~nm, showing a shoulder at about
520~nm.  The authors argued that such nanoparticles have spherical
geometry, however, we believe that it is  difficult assert because
the absorption peak is so wide in both cases that  any particular
optical signature due to geometry might be hidden by the  
distribution of sizes and shapes of the nanoparticles, and not only due to
the effects  in the change of the electron mean free path or in the
absorption energies of intra-band transitions, as it was argued~\cite{mafune1}. Unfortunately a direct comparison between our calculation and
their measurements is not possible since their samples  might have a
large distribution  of sizes, as well as a large variety of shapes.

\section{Conclusions}

 We use the discrete dipole approximation  to study the
main optical features of the extinction, absorption and scattering
efficiencies of nanoparticles of different sizes and shapes, made
out of silver and gold.  Here we have considered different
geometries:  spheres, ellipsoids, cubes, tetrahedra, cylinders and pyramids.
In most cases, we have clearly identified the main optical signature
associated to  each geometry. We  find,  as it might
have been expected, that the spectra are more complex as the particle has
less symmetry and /or has more  vertices. We have identified in the
spectra the main surface plasmon resonance associated to a dipolar
excitation, as well as other resonances due to high-multipolar excitations. 
 The physical origin of these high-multipolar excitations  %
might have two different sources, one due to the  shape and the
other due to the  size of the particle. In the case of silver
particles,  specific features in the optical spectra can be  %
associated to either geometry, size or material properties, making optical
spectroscopies  a very helpful tool for the characterization of
nanoparticles during and  after growth.  We also show that it
is not enough to  analyze the extinction efficiency  to
elucidate the size and shape of a nanoparticle,  but that is
absolutely necessary to do also an in-depth study of the absorption and
scattering efficiencies. 

 A direct comparison of our results with most of the available
experimental measurements of  the optical properties of suspended
nanoparticles  would require an averaging procedure over a wide
distribution of sizes and shapes.  But this averaging procedure
might smooth out the main relevant features of the spectra associated to the
size and shape of the nanoparticles. On the other hand,  it would be
very desirable to obtain optical spectra over samples with narrower
distributions of sizes and shapes.

\bigskip  We appreciate valuable discussions with Prof. Yves
Borensztein and Guillermo Ortiz. This work has been partly supported by
DGAPA-UNAM grant No.~IN104201 and by CONACyT grants~36651-E and~G32723-E.

\appendix  

\section{Surface effects}

One criteria to choose the number of dipoles or the  separation
distance between them is to avoid spurious surface effects.  We call
spurious surface effects to  those effects coming from the fact
that for  a given cubic lattice bounded by a specific particle shape,
it is possible to have a number of dipoles at the surface which can be
larger or comparable to the number of dipole within the particle.

Let us  assume that we have a sphere of volume $V$ which can be
discretized in N spherical entities with radius $d/2$, where $d$ is the
separation between dipoles. We have that $V\simeq Nv=4\pi a_{\mathrm{eff}%
}^{3}/3$, where $v$ is the volume ocuppied by each dipole $v=4\pi (d/2)^{3}/3
$, and $a_{\mathrm{eff}}$ is the effective radius of the particle. Now, we
want to know how many dipoles belong to the surface. Suppose we have $N_{s}$
dipoles at the surface which occupies a volume $N_{s}v$, 
\begin{equation}
N_{s}v=\frac{4\pi }{3}a_{\mathrm{eff}}^{3}-\frac{4\pi }{3}(a_{\mathrm{eff}%
}-d/2)^{3},
\end{equation}%
where we  find that 
\begin{equation}
N_{s}=N[1-\frac{1}{N}(N^{1/3}-1)^{3}].
\end{equation}%
Using this formula, we  have that for a total of $N=100$ dipoles,
about $52$ of them belong to the surface, for $N=1000$, about $271$ dipoles
belong to the surface, and so on, as shown in Table 1.

\newpage

\begin{table}[hbt]
\begin{center}
\caption{Number of surface dipoles according to Eq. A2}
\begin{tabular}{c|c|c}
\hline \hline
total dipoles $N$ & surface dipoles $N_s$ & $N/N_s (\%)$ \\ \hline
10$^2$ & 51.7 & 51.7 \\  
10$^3$ & 271.0 & 27.1 \\  
10$^4$ & 1328.8 & 13.2 \\  
10$^5$ & 6325.0 & 6.32 \\  
10$^6$ & 29701.0 & 2.97 \\ \hline \hline
\end{tabular}
\end{center}

\end{table}

\newpage

\begin{center}
FIGURE CAPTIONS
\end{center}

FIGURE 1: Optical coefficients for a silver nanosphere

FIGURE 2: Optical coefficients for a silver nanocube

FIGURE 3: Optical coefficients for a silver nanospheroid for an electric
field polarized along the minor axis

FIGURE 4: Optical coefficients for a silver nanospheroid for an electric
field polarized along the major axis

FIGURE 5: Optical coefficients for a silver nanocylinder for an incident
electromagnetic field along the symmetry axis

FIGURE 6: Optical coefficients for a silver tetrahedra for an incident
electromagnetic field perpendicular to the basis

FIGURE 7: Optical coefficients for a silver piramide for an incident
electromagnetic field perpendicular to the basis

FIGURE 8: Optical coefficients for a gold nanocube

\newpage

%



%
%

%


\begin{thebibliography}{99}
\bibitem{multipoles} Unfortunately, using DDA it is not a direct method to
assign a specific multipolar order to a given resonance structure. This is
because the near field around the particle is not well described in this
approximation. This has been pointed out before in Ref.~21.
\bibitem{drude} We performed calculations of absortion and scattering
effciencies with metal nanoparticles described by the Drude model using
different plasma frequencies ($\omega_p$) and relaxation times ($\tau$). In
all cases, independently of $\omega_p$ and $\tau$, we observed that for
large dielectric functions convergency problems are always present. On the
other hand, we have also performed calculations with semiconductor
nanoparticles and this problem of convergency is not observed. We have also
found that if the number of dipoles are increased dramatically from 65,000
to half a million, then the optical properties converges.
\end{thebibliography}
\end{document}